\documentstyle[aps,epsf,prc]{revtex}

\begin{document}
\draft
\title{Considerations on rescattering effects for threshold photo- and 
electro-production of $\pi^0$ on deuteron} 
\author{Michail P. Rekalo \footnote{Permanent address: 
\it National Science Center KFTI,
 310108 Kharkov, Ukraine}} 
\address{Middle East Technical University, Physics Department, Ankara 06531, 
Turkey}
\author{Egle Tomasi-Gustafsson \footnote{ E-mail:etomasi@cea.fr} }  
\address{DAPNIA/SPhN, C.E.A./Saclay,  91191 Gif-sur-Yvette Cedex, France}

\date{\today}
\maketitle
\begin{abstract}
We show that for the S-state $\pi^0$-production in processes 
$\gamma+d\to d+\pi^0$
and $e^-+d\to e^-+d+\pi^0$ the calculations of rescattering effects due to the  
transition:
$ \gamma+d\to p+p+\pi^-$ (or $n+n+\pi^+)\to d+\pi^0$ have to take carefully into account the Pauli principle. The large values for these effects predicted in the past  may 
result from the fact that 
the spin structure of the corresponding matrix element and the necessary 
antisymmetrization induced by the presence of identical protons (or neutrons) 
in 
the intermediate state was not taken into 
account accurately. One of the important consequences of these considerations 
is that $\pi^0$ photo- and 
electro-production on deuteron near 
threshold can bring direct information about elementary neutron amplitudes. 
\end{abstract}
\section{Introduction}
A considerable experimental and theoretical activity has been going on in the 
field of near 
threshold pion production in $\gamma N-$ and $eN-$ collisions. Recently, new 
results have been obtained for 
$\pi^0$-production on protons, $\gamma+p\to p+\pi^0$, using tagged 
photons 
\cite{Be97,Fu96}:
\begin{itemize}
\item the discovery of a unitary cusp in the energy dependence of the $E_{0+}$ 
amplitude for $\gamma+p\to p+\pi^0$, near the reaction threshold, 
\item the contradiction of the measured value of $E_{0+}(p\pi^0)$ with the 
predictions of "old" low energy theorems \cite{Kr54}.
\end{itemize}
The Chiral Perturbation Theory (ChPT) \cite{VB96} was very successful in the 
explanation of the properties of different S- and P-wave multipole amplitudes 
for $\gamma+p\to p+\pi^0$ in 
the near threshold region. The process of $\pi^0$ electro-production on 
protons \cite{We91,Di97,Me01}, $e^-+p\to e^-+p+\pi^0$, opens also new interesting 
possibilities due 
to the longitudinal polarization of the virtual photon and the non-trivial 
dependence of the multipole amplitudes on the momentum transfer 
squared, $Q^2$ from the initial to the scattered electron. But last data \cite{Me01} show a serious discrepancy with the calculation in the framework of heavy baryon ChPT \cite{Ber98}.

For a further test of different models, which can be applied to pion photo- and electro-production, the information about the amplitudes of the 
processes 
$\gamma+n\to 
n+\pi^0$ and $e^-+n\to e^-+n+\pi^0$ is essential. The 
deuteron processes 
$\gamma+d\to d+\pi^0$ and $e^-+d\to e^-+d+\pi^0$ seem well 
adapted for this aim, as, in 
impulse 
approximation (IA), the corresponding matrix elements are determined by the 
coherent sum of 
amplitudes 
for elementary processes $\gamma(\gamma^*)+p\to p+\pi^0$ and 
$\gamma(\gamma^*)+n\to 
n+\pi^0$ \cite{St90,Re91,Re93}, where $\gamma^*$ is the virtual photon. 
However, since, in the threshold region, these amplitudes are small in 
comparison with the amplitudes of 
$\pi^{\pm}$ photoproduction on nucleons,  rescattering effects, (RE) due to 
the following two-step processes: 
\begin{equation}
\gamma+d\to p+p+\pi^- ~(n+n+\pi^+)\to d+\pi^0
\end{equation}
can be, in principle, strongly competitive with the direct $\pi^0$-production 
from a single nucleon [12-19]. If it is the case, the reaction 
$\gamma+d\to d+\pi^0$ can not allow a 
direct and model independent extraction of the elementary neutron amplitudes.

A dedicated experiment \cite{Be98}, with a tagged photon beam, was devoted to 
measurements of the cross section and of the angular distribution  for the 
inclusive $\pi^0$-production in the $d(\gamma,\pi^0)X$ reaction in the near 
threshold region. Inelastic contributions, due to the deuteron 
disintegration, $\gamma+d\to \pi^0+n+p$ could not be resolved in the 
experiment, and their contribution was estimated through a theoretical model. 
It was shown 
that the electric dipole amplitude $E_{0+}$  for the  $\gamma+d\to 
\pi^0+d$ process (which is proportional to the amplitude referred later on as $g_e$) can be determined with good accuracy.  The data are 
characterized by a large
backward-forward asymmetry of the angular $\pi^0$-distribution. The deduced 
$E_{0+}$ amplitude,  has a negative sign, relative to the definite combination 
of P-wave amplitudes, which is  determined from a theory for $\gamma+N 
\to N+\pi$. The 
$E_{0+}$ amplitude for  $\gamma+d\to d+\pi^0$ is certainly sensitive 
to the the $E_{0+}$ amplitude for  $\gamma+n\to \pi^0+n$, but the 
quantitative determination of this last amplitude depends essentially on RE. The negative sign for $E_{0+}(\gamma d\to 
d\pi^0)$ was considered as a confirmation of the validity of ChPT predictions 
for  $\gamma+N\to N+\pi^0$, and of the important role of RE. 

Coherent $\pi^0$ threshold electro-production on the deuteron at $Q^2$=0.1 GeV$^2$ has been studied by the A1 collaboration at the Mainz Microtron MAMI, with $d\pi^0$-excitation energy up to $\Delta$ W = 4 MeV \cite{Ew01}. The longitudinal threshold amplitude, which was extracted by the Rosenbluth fit is smaller (in absolute value) by a factor two than the value predicted by ChPT calculations. The discrepancy at the level of the cross section is of one order of magnitude.

Note, in this connection, that the existing calculations of RE give very 
different quantitative predictions. These effects are very sensitive to many 
ingredients of the corresponding model, as, for example,  the short distance 
behavior of the deuteron wave 
function, the shape of the pion propagator in the intermediate 
state, the choice of the operator for the elementary process 
$\gamma+N\to N+\pi^0$, the procedure of integration (for example in \cite{St90} a six fold integration was done) etc. The accuracy of assumptions and 
simplifications in the theoretical calculations can  not be easily controlled. 
This may be the reason for which large discrepancies exist in the theoretical results.

Our main aim here is to demonstrate that for the threshold $S-$state $\pi^0$-meson 
production in processes 
$ \gamma+d\to d+\pi^0$ and $e^-+d\to e^-+d+\pi^0$ the 
application of Pauli 
principle and the conservation of P-parity and total angular momentum induces 
an cancellation of RE due to the reactions 
$ \gamma+d\to p+p+\pi^-$ (or $n+n+\pi^+)\to d+\pi^0$.

This paper is organized as follows. In Section II we establish the spin structure of the threshold amplitudes for $\gamma+d\to d+\pi^0$ and analyze the most simple polarization phenomena for this process. The importance of Pauli principle for RE, due to the processes $\gamma+d\to p+p+\pi^-(n+n+\pi^+)\to d+\pi^0$, is demonstrated in Section III. The properties of the dispersive part of RE are described in Section IV. The description of the process $\gamma+d\to p+p+\pi^-$ in the framework of impulse approximation is presented in Section V. In Section VI we consider the general kinematics of the threshold inelastic $\pi$-meson photoproduction, $ \gamma+d\to N+N+\pi$ for charged and neutral pions. Finally, in Section VII we consider the 'scalar' deuteron photoproduction, $\gamma+d\to (n+p)_s+\pi^0$, (i.e. the $np$-system is in a singlet state) where RE must be important.

\section{Spin structure of threshold amplitudes for $\gamma+d\to 
d+\pi^0$}

Let us consider, firstly, the simplest process of pion photoproduction on the 
deuteron, $\gamma+d\to d+\pi^0$, in the threshold region, 
where the main contributions are due to $\pi^0$ production in S- and P-states. 
For S-state 
production (i.e. with ${\cal J}^P=1^-$, where ${\cal J}$ is the total 
angular momentum of the produced $d\pi^0$-system and $P$ is its $P$-parity) the 
conservation of $P$-parity and total angular 
momentum 
allows two multipole transitions, $E1$ and $M2\to  {\cal J}^P=1^-$, 
with the 
following 
parametrization of the spin structure of corresponding matrix element:
\begin{equation}
{\cal M}(\gamma d \to d\pi^0)=g_e\vec e\cdot\vec D_1\times \vec 
D^*_2+g_m
(\vec e\times \vec k\cdot\vec D_1~\vec k\cdot \vec D_2^*+\vec e\times \vec 
k\cdot\vec D_2^*~\vec k\cdot \vec D_1),
\end{equation}
where $\vec e$ is the photon polarization vector, $\vec D_1$ and $\vec D_2$ 
are the axial vectors of 
deuteron polarization (in initial and final states), $\vec k$ is the unit 
vector along the three-momentum of 
photon in the center of mass (CMS) of the considered reaction, $g_e$ and $g_m$ are the multipole amplitudes, describing the 
$E1-$ and $M2-$photon absorption with the S-state $\pi^0$-production. In the 
general case these 
amplitudes are complex functions of the photon energy, $E_{\gamma}$, but 
due to the T-invariance of hadron 
electrodynamics \cite{Ch66}, the relative phase of multipole amplitudes must be equal to $0$ 
or $\pi$.

The differential cross section for the process $\gamma+d\to d+\pi^0$
is expressed in terms of the multipole amplitudes $g_e$ and $g_m$, in a particular normalization, as:
$$
\left (\displaystyle\frac {d\sigma}{d \Omega}\right )_0=\frac{2}{3}\left ( 
|g_e|^2+|g_m|^2\right ).
$$
For S-wave pion production, among all possible one-spin polarization observables,  only the tensor analyzing 
power ${\cal A}$ (in $\gamma+\vec d\to d+\pi^0$) does not vanish:
\begin{equation}
\displaystyle\frac {d\sigma}{d \Omega}(\gamma \vec d)=
\left ( \displaystyle\frac {d\sigma}{d \Omega}
\right )_0
\left [ 1+(Q_{ab}k_ak_b){\cal A} \right ],
\end{equation}
where $Q_{ab}$ is the tensor polarization of deuteron target. The corresponding density matrix can be written as:
$$
\overline{D_{1a}D^*_{1b}}=\frac{1}{3}(\delta_{ab}-\frac{3}{2}i\epsilon_{abc}S_
c-Q_{ab}),~Q_{ab}=Q_{ba},~Q_{aa}=0,$$
where $\vec S$ is the vector deuteron polarization.
Averaging over the photon polarizations and summing over the final deuteron polarizations, one finds:
\begin{equation}
{\cal A}=-\displaystyle\frac{1}{4}\displaystyle\frac{|g_e|^2+|g_m|^2+6{\cal 
R}e~g_eg_m^*}{|g_e|^2+|g_m|^2}=- \left ( \displaystyle\frac{1}{4}\pm \displaystyle\frac{3}{2}\displaystyle\frac{r}{1+r^2}\right ),
\end{equation}
where we defined the ratio $r=|g_m|/|g_e|$. The sign $\pm$ corresponds to the two possible relative signs of the amplitudes $g_e$ and $g_m$. The behavior of the asymmetry ${\cal A}$ as a function of $r$ is shown in Fig. 1. One can see the large sensitivity of ${\cal A}$ to the relative sign and to the ratio of the $g_e$ and $g_m$ amplitudes, even at small $r$. 

The complete experiment, i.e. the full determination 
of the amplitudes $g_e$ and $g_m$, is realized, at threshold, through the measurement of two observables only, 
the differential cross section and the tensor analyzing power:
$$
|g_e+g_m|^2=(1-2{\cal A})\left (\displaystyle\frac {d\sigma}{d \Omega}\right 
)_0,~~
|g_e-g_m|^2=2(1+{\cal A})\left (\displaystyle\frac {d\sigma}{d \Omega}\right 
)_0.
$$
The interference contribution, which is sensitive to the small magnetic 
amplitude $g_m$, can be determined through the following formula:
$$g_eg_m=-\displaystyle\frac {1}{4}(1+4{\cal A})\left (\displaystyle\frac{d\sigma}{d \Omega}\right )_0.
$$
Note that all the polarization phenomena for the process  $\gamma+d\to 
d+\pi^0$ near threshold can be predicted in terms of the tensor analyzing power and of the differential cross section. To demonstrate this, let us consider, as an example, the collision of polarized photons with a 
polarized deuteron target. In case of a linearly polarized photon beam, we can 
define the following asymmetry:
$$
\Sigma=\displaystyle\frac{d\sigma(e_x) -d\sigma (e_y) }
{d\sigma(e_x)+ d\sigma(e_y)}$$
where $e_x$ and $e_y$ are the components of the photon polarization vector in 
a coordinate system with the $z-$axis along the 3-vector $\vec k$. 

The asymmetry $\Sigma$ can be written as a function of the multipole amplitudes $g_e$ and $g_m$, and of the tensor polarizations as:
$$
\Sigma=\displaystyle\frac{(Q_{xx}-Q_{yy})|g_e-g_m|^2}
{(|g_e|^2+|g_m|^2)(4-Q_{zz})-6g_eg_mQ_{zz}},
$$
so that:
$$
\Sigma\displaystyle\frac{d\sigma}{d \Omega}(\gamma\vec 
d)=2\left (Q_{xx}-Q_{yy}\right )(1+{\cal A})
\left(\displaystyle\frac{d\sigma}{d \Omega}\right )_0.
$$
In IA, for $\gamma+d\to d+\pi^0$, (Fig. 2), the 
amplitude $g_e$, which is 
generated by the S-wave component of the deuteron wave function (at relatively 
small internal momentum), is proportional to the sum of $E_{0+}$ amplitudes 
for $\gamma+p\to p+\pi^0$ and $\gamma+n\to n+\pi^0$-processes:
\begin{equation}
g_e=F_s(t)\left [ E_{0+}^{p\pi^0}+E_{0+}^{n\pi^0}\right ],
\label{eq:geia}
\end{equation}
where $F_s(t)$ is the S-wave deuteron form factor at $t=-E_{th}^2$ and  
$E_{th}$ is the photon threshold energy (in CMS). 
The amplitude $g_m$ of magnetic quadrupole absorption has not an analogue for the elementary processes $\gamma+N\to N+\pi$, at 
threshold. In the framework of IA, it can not be derived from the spin structure $\vec\sigma\cdot\vec e$  for the threshold amplitude of the elementary process 
$\gamma+N\to N+\pi$. Therefore 
the amplitude $g_m$ is very sensitive to the details of the reaction mechanism, in $\gamma+d\to d+\pi^0$. The realization of the complete 
experiment, as indicated above, would be very interesting, in this respect.
\section{Cancellation of rescattering contributions}

The impulse approximation is only one of the ingredients in the 
analysis of the process $\gamma(\gamma^*)+d\to d+\pi^0$. 
Let's discuss now RE.

It is well known (\cite{Ko77,Bo78} and refs. herein) that the $E_{0+}$ amplitude for the charged pion production 
on nucleons is larger (in 
absolute 
value) than the corresponding amplitude for neutral pion production: $|E_{0+}^{n\pi^+}|\simeq 20 |E_{0+}^{p\pi^0}|$. This fact is generally accepted as the underlying reason for the manifestation of RE due to reactions (1). In this case, a model independent information about the elementary amplitude of $\pi^0$-production on neutron, $\gamma+n\to n+\pi^0$ can not be derived from the study of the reaction $\gamma+d\to d+\pi^0$. Previous 
calculations \cite{Ko77,Bo78,Fa80} have 
shown that RE, which involve intermediate charged pions, essentially change 
the predictions of the impulse approximation.

We show here that RE due to reactions (1), cancel out for the threshold 
amplitudes, if one takes into account 
the spin structure of the corresponding transitions which are allowed by the 
Pauli principle and the conservation of angular momentum and P-parity. As a result,  we will prove that the state with ${\cal J}^P=1^-$ is forbidden for the intermediate
$pp\pi^-$(and $nn\pi^+$)- system, if $\ell_1=\ell_2=0$ (threshold conditions), 
where $\ell_1$ is the 
orbital angular momentum for the $pp-$system, and $\ell_2$ is the pion orbital 
momentum relative to the pp-system (Fig. 3). Therefore reactions (1) can not occur in threshold regime.

For the $pp-$ system with $\ell_1=0$, only the singlet state is allowed. Therefore, at threshold 
of 
$\gamma+d\to p+p+\pi^-$, ${\cal J}^P$ takes the value $0^-$ (instead of ${\cal J}^P=1^-$ for the threshold $d+\pi^0$-system).
This is illustrated also in the corresponding Feynman diagrams (Fig. 
4) , where at threshold, we have for the three-momentum of the protons: $\vec 
p_1=\vec p_2=0$, with evident cancellation of the two contributions.

This result is valid for any parametrization of 
the deuteron wave function and relative value of the amplitudes for the 
different pion production processes. It is also correct  for coherent 
$\pi^0$ electro-production on deuteron, $e+d\to e+d+\pi^0$, at any value of momentum transfer square $Q^2$,  in the 
space-like region and for any polarization (transversal and longitudinal) of 
virtual photons. The main assumption, done here, is that both nucleons in 
the intermediate state are on mass shell, so that the $pp$-system, with 
$\ell_1=0$ has positive P-parity. Off-shell protons would have an antinucleon 
component with negative P-parity. In principle, these intermediate 
configurations can contribute, but they have been neglected in all previously 
quoted calculations of RE, which were done in framework of non-relativistic 
approach.

The  cancellation of RE (for the imaginary part of the threshold amplitudes) in the $S-$ state, for $\gamma+d\to d+\pi^0$ at 
threshold is a rigorous general result, which has to be verified by any model 
calculation. But, technically, this can be a  difficult problem.
To show this, let us consider the standard procedure of RE calculation, with 
a single 
$\pi N$-scattering. To satisfy the Pauli principle, (for the intermediate $\pi 
NN$-state, with two identical nucleons), it 
is necessary to add to the usual diagram (where a pion, photoproduced on one 
nucleon, is scattered by another one, Fig. 5a), the diagram (5b), 
where the pion is scattered by the same nucleon.

Only the sum of (a)+(b) contributions, calculated with the same vertexes, 
satisfies the Pauli principle, resulting in a compensation of RE. But 
the (5b) contribution, is a particular part of the amplitude for $\gamma+d\to d+\pi^0$, calculated in IA 
(Fig. 6). So, to avoid a double counting, in model calculations it is necessary 
to subtract from the full IA $\gamma+ N\to N+\pi^0$ amplitude, the 
important part due to pion rescattering on the same nucleon: 
$\gamma+p\to n+\pi^+\to p+\pi^-$. This 
means that the IA amplitude for $\gamma+d\to d+\pi^0$ must be 
calculated with a {\it renormalized} amplitude for $\gamma +p\to 
p+\pi^0$, (Fig. 7), not with the standard one. 
This renormalization procedure is nontrivial and looks like a numerical 
artifact, but the 
calculations of RE, due to (5a) only, violate the Pauli principle and result 
in large RE for $\gamma+d\to d+\pi^0$.

Therefore, the most delicate problem in calculating the contributions (5a) 
and (5b), is to satisfy the Pauli principle and to avoid  double 
counting, which can induce large, non-physical RE.
Let us reanalyze, at the light of the previous discussion, the available experimental data. The value of the threshold amplitude, extracted from deuteron photoproduction data \cite{Be98}, is $E_d=(-1.45\pm 0.09)\times 10^{-3}/m_{\pi}$. More exactly, the experimental observable is the cross section, which is proportional to the amplitude squared. The minus sign has been attributed in order to be consistent with the ChPT predictions \cite{Bea97}. In the framework of IA, \ref{eq:geia}, if we assume that RE are absent, it is straightforward to extract the neutron amplitude. In Table I we report the values of the neutron amplitude for the two possible signs of the deuteron amplitude. We remark that the result obtained for $E_{0+}^{n\pi^0}$, using a positive value for $E_d$ and the experimental value of 
$E_{0+}^{p\pi^0}=-1.13$ \cite{Be97}, is not far from the ChPT prediction.

The  cancellation of RE's  at the threshold of the process 
$\gamma+d\to d+\pi^0$, can explain naturally and in a model 
independent way, the absence of unitary cusp in the energy dependence of the 
$E_{0+}$-amplitude for this process recently experimentally observed 
\cite{Be98}. This cusp is present on $\pi^0$-photoproduction on the 
nucleon, due to the $\gamma+p\to n+\pi^+\to p+\pi^0$ 
rescattering and they have been observed on a proton target \cite{Be97,Fu96}, 
but the Pauli principle forbids the corresponding intermediate 
states in (1) (in case of a deuteron target). Therefore the absence of cusp can 
be considered as an 
experimental 
evidence of the cancellation of RE in threshold $\pi^0$ 
photoproduction on deuteron. 

Note also that the experimental data about coherent $\pi^0$  electro-production on the deuteron at $Q^2$=0.1 GeV$^2$, do not show either any evidence of the corresponding cusp at 2.2 MeV above the $\pi^0$-threshold \cite{Ew01}. This cusp should be present if RE, due to (1) were important.

Summarizing the previous discussion, the most crucial points in the evaluation 
of  RE, especially in numerical calculations, in the 
near threshold region for $\gamma+d\to d+\pi^0$ are the following:
\begin{itemize}
\item cancellation of S-wave contributions (independently for $\pi^- pp$ and 
$\pi^+ nn$ intermediate states), 
which must be done analytically, in exact form;
\item estimation of the relative role of other possible {\it non S-wave } 
contributions to the RE.
\end{itemize}

\section{Dispersive contributions to RE}
The above mentioned result  does not mean that all rescattering effects cancel in the near threshold region for the process $\gamma(\gamma^*)+d\to d+\pi^0$. More precisely it is correct for the imaginary part of the two threshold amplitudes,
$g_e$ and $g_m$, for $\gamma(\gamma^*)+d\to d+\pi^0$, with $NN\pi^\pm$ intermediate states, where all these three particles, being in  relative S-states, are on mass and energy shell. These imaginary parts vanish, due to the Pauli principle and P-parity conservation. Such cancellations explain naturally the absence of cusp in the energy dependence of the threshold amplitudes for $\gamma+d\to d+\pi^0$ in the corresponding experimental data, whereas this cusp is present in the energy dependence of the $E_{0+}( \gamma p\to p+\pi^0)$ amplitude, due to the unitarity chain:  $\gamma p\to n+\pi^+\to p+\pi^0$.

But what about the real (dispersive) part of the amplitudes $g_e$ and $g_m$, corresponding to ${\cal J}^P=1^-$? These quantities are determined by definite dispersion integrals, from the corresponding imaginary parts - over the photon energy (and over the internal momenta of the $NN\pi$-system) from threshold to infinity. However, far from threshold, ${\cal I}m g_{e,m}$ contains different contributions, with higher values of the orbital momenta $\ell_1$ and $\ell_2$. The P-parity conservation and the Pauli principle have to be taken into account for the analysis of these contributions. For example, the values $\ell_1$ and $\ell_2$ must be even for the singlet $NN-$system, therefore the lowest $NN\pi^\pm$-intermediate state is characterized by $\ell_1=\ell_2=2$. The next states have $\ell_1=\ell_2=4$, etc. Such contributions to ${\cal I}m g_{e,m}$ can not be generated by the threshold $\vec\sigma\cdot \vec e$-operator for the elementary process $\gamma+N\to N+\pi$ and start to appear far from the threshold of the process $\gamma+d\to d+\pi^0$.

The selection rules ( due to P-parity conservation and Pauli principle) allow intermediate  $NN\pi^\pm$-states where the nucleons are in a triplet state with odd values of $\ell_1$ and $\ell_2$. The first of these contributions to ${\cal I}m g_{e,m}$has the following quantum numbers: $\ell_1=\ell_2=1$, $S_{NN}=1$, and thereafter $\ell_1=\ell_2=3$, $S_{NN}=1$, etc.

So for all these intermediate states with nonzero values of $\ell_1$ and $\ell_2$, the $\pi N$-elastic scattering, (which is the next step in the rescattering chain, see, for example, Fig. 9), can not occur in the $S$-state, as it was often assumed in the estimations of RE.

Again we must stress that these states have a 'non-threshold' nature, because they can not be generated by the threshold operator $\vec\sigma\cdot \vec e$. In the previous theoretical considerations \cite{Ko77,Bo78,Fa80} namely this operator was responsible for the large RE in the $\gamma+d\to d+\pi^0$-process near threshold. The argument to justify such approximation was based on the inequality $|E_{0+}^{\gamma p\to n\pi^+}|\simeq 20 |E_{0+}^{\gamma p\to p\pi^0}|$. But as we showed above, large RE near threshold can be induced by the electric dipole contribution only by contradicting the Pauli principle. We showed that RE in the threshold region are canceled in the imaginary parts of $g_e$ and $g_m$. We did not calculate the dispersive part of RE, where, evidently, it will be necessary to consider not only 
$NN\pi^\pm$ -states, but $NN\pi^0$ states - with neutral pions- as well.

In a similar way, RE can be analyzed not only for the S-wave production in $\gamma+d\to d+\pi^0$, but for P-wave production also, and for higher waves as well. The P-parity conservation and the Pauli principle will be equally important for the analysis of possible RE contributions to the corresponding imaginary parts of multipole amplitudes, showing the cancellation of many contributions. Therefore, probably such way -through the calculation of multipole amplitudes in two steps - finding the imaginary part, firstly, and then calculating the dispersive part, will be the most effective way to perform a correct evaluation of RE.

\section{Attempt of multipole analysis}
An threshold, i.e. when $\ell_1=\ell_2=0$, for the process $\gamma+d\to p+p+\pi^-~(n+n+\pi^+)$ we showed that RE cancel 
out. Other values of $\ell_1$ and $\ell_2$ are, in principle, 
possible, but their contribution can not be 
large, in the threshold region, due to centrifugal considerations. Let us consider $\ell_1=\ell_2=1$ (Fig. 3), 
which is the next allowed 
possibility to obtain ${\cal J}^P=1^-$, the threshold value. In order to generate such states, it is necessary to have a particular mechanism of RE. In the framework of the existing  analysis of RE, 
based on the standard structure of the 
$\gamma+ N\to N+\pi$ near-threshold amplitude, $E_{0+}\vec\sigma\cdot\vec e$, it 
is  possible to show that  two P-waves for the $\pi NN$- 
intermediate state are not allowed.

The matrix element for 
$\gamma +d\to p+p+\pi^-$ in the case of ${\cal J}^P=1^-$, with 
$\ell_1=\ell_2=1$  is proportional to the product of two small three-momenta: 
$\vec p$ (proton) and 
$\vec q$ (pion). Let us parametrize these contributions in a model independent 
way. The 
conservation of the total angular momentum and the P-parity allows the 
following multipole 
transitions for $\gamma+d\to \pi^-+p+p~( {\cal 
J}^P=1^-,~\ell_1=\ell_2=1)$: $E1$ and 
$M2\to~j=0,~1$ and $2$, where $j$ is the total angular momentum of the 
produced $pp-$system. Having $\ell_1=1$, such system has to be in triplet 
state, so $j=\vec 1+\vec 1=0,~1,~2$.

The spin structure of these transitions is:
$$\chi_2^{\dagger}\vec\sigma\cdot\vec p\sigma_y\tilde{\chi_1}^{\dagger}~\vec 
q\cdot\vec e\times\vec D,~~E1\to j=0,$$
$$\chi_2^{\dagger}\left(\vec\sigma\cdot\vec q~\vec e\times\vec D \cdot\vec 
p-\vec\sigma\cdot\vec 
e\times\vec D~\vec p \cdot\vec q\right 
)\sigma_y\tilde{\chi_1}^{\dagger},~~E1\to j=1,$$
$$\chi_2^{\dagger}\left(\vec\sigma\cdot\vec e\times\vec D~\vec p\cdot\vec 
q+\vec\sigma\cdot\vec q~
\vec p \cdot\vec e\times\vec D-\frac{2}{3}\vec\sigma\cdot\vec p~\vec 
q\cdot\vec e\times\vec 
D\right 
)\sigma_y\tilde{\chi_1}^{\dagger},~~E1\to j=2,$$
$$\chi_2^{\dagger}~\vec\sigma\cdot\vec p~\sigma_y\tilde{\chi_1}^{\dagger}
\left (\vec e\times\vec 
k\right )\times\vec D\cdot\vec q,~~M2\to j=0,$$
$$\chi_2^{\dagger} \left(\vec\sigma\cdot\vec q~p_a-\sigma_a\vec p\cdot\vec 
q~\right )
\sigma_y\tilde{\chi_1}^{\dagger}\left [ (\vec e\times\vec k )_a \vec k\cdot 
\vec D+k_a\vec 
e\times \vec k\cdot\vec D\right ],~~M2\to j=1,$$
$$\chi_2^{\dagger} \left(\sigma_a\vec p\cdot\vec q+ p_a\vec\sigma\cdot\vec 
q-\frac{2}{3}q_a\vec\sigma\cdot\vec p\right 
)\sigma_y\tilde{\chi_1}^{\dagger}\left([\vec 
e\times\vec k]_a \vec k\cdot \vec D+k_a\vec e\times\vec k \cdot \vec D\right 
),~~~M2\to j=2.$$
Such spin structure can not be generated by the two 
diagrams (Fig. 4), which 
are typically used in the standard calculations of RE, as the sum of these diagrams is proportional to:
$$
\chi_2^{\dagger}\left [\vec e\cdot\vec D (u_s+u'_s)+\vec e\cdot \vec p~\vec 
p\cdot\vec 
D(u_d+u'_d)\right . 
$$
\begin{equation}
\left . +i\vec\sigma\cdot\vec e\times\vec D(u_s-u'_s)+\vec p\cdot\vec 
D~\vec\sigma\cdot\vec 
e\times\vec p(u_d-u'_d)\right ]\sigma_y\tilde{\chi_1}^{\dagger}, 
\label{eq:first}
\end{equation}
where $u_s$ and $u_d$ are two possible S- and D-components of the nonrelativistic deuteron wave function which depend  on $|\vec 
k+\vec p|^2$, while $u'_s$ and $u'_d$ depend on $|\vec k-\vec p|^2$. 

Therefore, in threshold region, where $\vec p\simeq 0$, $u_s-u'_s\to 0$, 
and $u_d-u'_d\to 0$, this 
mechanism can induce the following transitions for the $\pi^- pp$ system: 
${\cal J}^P=0^-$ with $\ell_1=\ell_2=0$ and ${\cal J}^P=2^-$ with $\ell_1=\ell_2=0$ (both protons are in a 
singlet state). But ${\cal J}^P=0^-$ is not a possible configuration for the 
reaction $\gamma+d \to d+\pi^0$ at threshold (as it was proved above),  and  ${\cal J}^P=2^-$ 
corresponds to the D-wave of $\pi^0$, with evidently 
small amplitudes. The $NN$- final interaction (Fig. 8a) can not transform a 
singlet $pp$-system to 
a triplet one, and $\pi N$ rescattering (Fig. 8b) can not re-arrange the 
threshold spin structure 
of the matrix element for $\gamma+d \to p+p+\pi^-$ in the framework of 
the considered mechanism.

The possible triplet contributions to the matrix element (\ref{eq:first}), 
may appear only far 
from threshold, where $u_s\ne u'_s $ and  $u_d\ne u'_d $: in this case the 
states with $\ell_1=1$ and $3$ are possible, but then the P-parity of this 
channel is positive, because $\ell_2=0$, and, again, it is incompatible with ${\cal J}^P=1^-$ (which characterizes the threshold conditions for $\gamma+d \to d+\pi^0$).

For a more general analysis, it is possible to take into account the full spin 
structure of the elementary process $\gamma +N\to N+\pi$ in the 
following form:
$$\vec\sigma\cdot\vec e~f_1+i\vec e\cdot\vec k \times\vec q~f_2+\vec e\cdot\vec q~ 
\vec\sigma\cdot\vec 
k~f_3+\vec e\cdot\vec q~\vec\sigma\cdot\vec q~f_4.$$
One can see that in threshold conditions for the $pp-$system, at this vertex  
all configurations for any value of $\ell_2$ are allowed, but there is a 
restriction on $\ell_1$: only singlet $pp-$states with $\ell_1=0$ or $\ell_1=2$
are permitted. 
And only one configuration with a combination of $\ell_2=2$ and $\ell_1=2$ can 
produce ${\cal J}^P=1^-$. 
However this is a small contribution, which results from the 
multiplication of at least 3 small factors:
$D-wave~of~deuteron\bigotimes$ $D-wave~in~ \gamma+N\to 
N+\pi$$\bigotimes ~D-wave~in~ 
threshold ~N+N-system$.

Note that  P-wave in $\gamma +N\to N+\pi^+$ combined with the effect due 
to $u_s-u'_s$ (or 
$u_d-u'_d$) can result in ${\cal J}^P=1^-$. But we showed that this effect is 
small and, moreover, 
it is not related to the fact that $E_{0+}^{n\pi^+}\simeq 20$~$E_{0+}^{p\pi^0}$, 
which is commonly 
taken as an evidence of large RE. The standard mechanism of 
pion rescattering in 
the reactions $\gamma+d\to p+p+\pi^-\to d+\pi^0$, cannot produce 
large multipole 
amplitudes with  ${\cal J}^P=1^-$.

A similar analysis of RE holds also for threshold $\pi^0$ electro-production, 
$e+d\to e+d+\pi^0$. Let us mention in this respect, that the threshold matrix 
element for 
$\gamma^*+d\to d+\pi^0$ contains the contributions of the 3 following 
multipole transitions: $E1_t$, $E1_{\ell}$ and $M2\to {\cal J}^P=1^-$, 
where the indexes $t$ and $\ell$ correspond to 
the absorption of a virtual photon with transversal and longitudinal 
polarization. Therefore the matrix element for the process 
$\gamma^*+d\to d+\pi^0$, corresponding to S-state $\pi^0$-production 
has the following expression \cite{Re97}:
\begin{eqnarray*}
{\cal M}(\gamma^*+d\to d+\pi^0)&=&g_t(k^2)(\vec e\cdot \vec D_1\times 
\vec D_2^*-\vec 
e\cdot\vec k~\vec k\cdot\vec D_1\times\vec {D^*_2})+\\
&&+g_{\ell}(k^2)\vec e\cdot\vec k~\vec k\cdot\vec {D_1}\times\vec{D_2^*}+\\
&&+g_m(k^2)(\vec e\times\vec k\cdot\vec D_1~\vec k\cdot \vec{D_2^*}+\vec 
e\times\vec k\cdot\vec {D_2^*}\vec k\cdot\vec D_1),
\end{eqnarray*}
where $g_t$,  $g_{\ell}$ and $g_m$ are the corresponding form 
factors, depending, in general, on two kinematical variables, $k^2$ and $s$. In 
any case they can be 
considered as the inelastic threshold electromagnetic form factors for the 
S-state $\pi^0$-production in the process 
$\gamma^*+d\to d+\pi^0$. This parametrization of the matrix element 
(which is equivalent to the corresponding description of the elastic $ed-$ 
scattering, in terms of three well-known form factors) is suitable for the description of polarization phenomena for $e^-+d\to e^-+d+\pi^0$, near threshold. Consequently, a Rosenbluth fit for 
$e^-+d\to e^-+d+\pi^0$ (with unpolarized 
particles) allows us to find two quadratic combinations of form factors, 
namely $\sigma_L\simeq |g_{\ell}|^2$ and $\sigma_T\simeq|g_t|^2+|g_m|^2$. In order to separate the 
$g_t$ and $g_m$ contributions, the measurement of the tensor analyzing power 
(or the final deuteron tensor polarization) is necessary, as in the case of elastic $ed$-scattering. 
\section{Spin structure of threshold amplitudes for $\gamma+d\to 
N+N+\pi$}
In case of neutral pion production in the intermediate state, 
 $\gamma+d\to n+p+\pi^0\to d+\pi^0$, RE can contribute to threshold $\pi^0$ production, $\gamma+d\to d+\pi^0$ through the triplet $np-$intermediate state,  due to the non identity of $n$ and $p$, but these effects are small. This follows from the fact that the spin structure of the threshold amplitudes for the 
processes $\gamma+d\to p+p+\pi^-$ and $\gamma+d\to n+p+\pi^0$ are different.

In this connection we can mention that the thresholds for $\gamma +d\to 
p+p+\pi^-$ and 
$\gamma+d\to n+n+\pi^+$ processes are higher in comparison with $ \gamma 
+d\to d+\pi^0$:
$$E_{\gamma}(pp\pi^-)= 145.8~ \mbox{MeV},$$
$$E_{\gamma}(nn\pi^+)= 148.5~ \mbox{MeV},$$
$$E_{\gamma}(d\pi^0)~= 139.8~\mbox{MeV}.$$

\noindent\underline{$\gamma+d\to p+p+\pi^-$}: the spin structure of 
this amplitude is mainly driven by the Pauli principle. At threshold the 
single allowed multipole transition 
$E1\to {\cal J}^P=0^-$ is described by  the following matrix element:
$${\cal M}(pp\pi^-)=f_0\vec e\cdot\vec 
D_1~\chi_2^{\dagger}\sigma_y\tilde{\chi_1}^{\dagger}.$$
The amplitude $f_0$ describes the absorption of electric dipole photons and 
$\chi_1$ and $\chi_2$ are the 2-component spinors of the final nucleons. 

\vspace*{.2true cm}
\noindent\underline{$\gamma+d\to p+n+\pi^0$ }: at threshold we have 
three independent multipole transitions: $E1\to {\cal J}^P=0^-$ (singlet $np-$production as in the case of $\gamma+d\to p+p+\pi^-$), and the following two multipole transitions: $E1$ and $M2~\to$ $ {\cal J}^P=1^-$, - with triplet 
$np-$production - described by the  matrix element:
$${\cal M}(np\pi^0)=\chi_2^{\dagger}\left [\vec\sigma\cdot\vec e\times\vec D_1 
f_e+(\vec 
e\times\vec k\cdot\vec\sigma~\vec k\cdot\vec{D_1}+\vec\sigma\cdot\vec k~\vec 
e\times\vec 
k\cdot\vec D_1)f_m\right ] \sigma_y\tilde{\chi_1}^{\dagger},
$$
where $f_e$ and $f_m$ are the corresponding multipole amplitudes. From the generalized Pauli principle, we can conclude that these amplitudes are isovector.

All polarization phenomena in 
$\gamma+d\to p+p+\pi^-$ can be exactly predicted, in a model independent 
way due to the presence of a single threshold amplitude. For 
example, the dependence of the cross section from the deuteron tensor 
polarization can be written as:
$$\displaystyle\frac{d\sigma}{d\omega}(\gamma \vec d\to pp\pi^-)=\left 
(\displaystyle\frac{d\sigma}{d\omega}\right 
)_0\left[1+\frac{1}{2}Q_{ab}k_ak_b\right ],$$
where $d\omega$ is the element of the space volume for the 3-particle 
final state.

The presence of two threshold amplitudes, $g_e$ and $g_m$, for 
$\gamma+ d\to p+n+\pi^0$ 
results in different sign and value for the deuteron analyzing power:
$$\displaystyle\frac{d\sigma}{d\omega}(\gamma \vec d\to np\pi^0)=\left 
(\displaystyle\frac{d\sigma}{d\omega}\right 
)_0\left[1-\frac{1}{4}Q_{ab}k_ak_b\right ].$$
At threshold, all other one-spin polarization observables for the processes $\gamma+d\to N+N+\pi$  vanish. 

\section{The process $\gamma+d\to d_s+\pi^0$}

The situation with RE in the semi-coherent 
process $\gamma+d\to p+n+\pi^0\to d_s+\pi^0$, with 
production of a 'scalar' 
deuteron, $d_s$, is very different, in comparison with the reaction 
$\gamma+d\to d+\pi^0$.  For the process 
$\gamma+ d\to d_s+\pi^0$, at threshold,  ${\cal J}^P$ takes the value $0^-$ and the 
conservation of the total angular momentum and P-parity allows the following 
intermediate processes (where the intermediate $pp\pi^-$ $(nn \pi^+)$ have also ${\cal J}^P=0^-$):
$$ \gamma+ d\to p+p+\pi^- (n+n+\pi^+)\to d_s+\pi^0.$$
Both rescattering contributions (Fig. 9) have the same sign: this induces a 
coherent increasing, instead of a cancellation (as in the case $\gamma+ 
d\to d+\pi^0$). 

Another interesting property of the considered process is connected with the isotopic structure of the corresponding amplitude. The production of $d_s$, with unit value of isotopic 
spin in the reaction $\gamma+d\to d_s+\pi^0$, is determined by the 
absorption of isoscalar photon (instead of isovector, for the process 
$\gamma+ d\to d+\pi^0$). In IA, (Fig. 10), the corresponding amplitudes 
is 
proportional to the difference of the elementary amplitudes, i.e.:
$$ F(\gamma d\to d_s\pi^0)\simeq (E_{0+}^{p\pi^0}-E_{0+}^{n\pi^0}).$$
As the elementary amplitudes, have different signs, one can 
deduce that cross 
section for the process 
$\gamma+d\to d_s+\pi^0$ is much larger in comparison with 
$\gamma+ d\to d+\pi^0$. Therefore the ratio of these cross sections 
has to be very sensitive to the model chosen to determine $E_{0+}^{p\pi^0}$. 
For example, in the framework of ChPT, with $E_{0+}^{p\pi^0}=-1.16$ and 
$E_{0+}^{n\pi^0}=2.13$ (in units $10^{-3}/m_{\pi}$), one can find:
$$\displaystyle\frac{|F(\gamma d\to d_s\pi^0)|^2}{|F(\gamma 
d\to d\pi^0)|^2}=\left |\displaystyle\frac{2.13+1.16}{2.13-1.16}\right 
|^2\simeq 10.$$
In case of the dispersion relations approach (DR)\cite{Ha97}, with 
$E_{0+}^{p\pi^0}=-1.22$ and $E_{0+}^{n\pi^0}=1.19$, this ratio is much larger:
$$\displaystyle\frac{|F(\gamma d\to d_s\pi^0)|^2}
{|F(\gamma d\to d\pi^0)|^2}=\left 
|\displaystyle\frac{1.19+1.22}{1.19-1.22}\right 
|^2\simeq 6450.$$ 
This consideration shows that inelastic pion production, in 
the near threshold region, can be very large for the process 
$\gamma+d\to d_s+\pi^0$ and  should be determined 
experimentally, too.
A similar situation occurs for $\eta$ photoproduction on the deuteron. The first 
experimental study 
of this process was done more than 25 years ago \cite{An69}, but only 
recently it has been shown \cite{Kr95,Ho97} that the cross section of the 
coherent  process $\gamma+d\to d+\eta$ is smaller than the incoherent 
$\eta$-photoproduction through $\gamma+d\to p+n+\eta$. 
\section{Conclusions}

We have shown  that RE for the processes $\gamma+d\to d+\pi^0$ and 
$e+d\to e+d+\pi^0$ , due to the intermediate processes $\gamma+ d\to 
p+p+\pi^-(or~n+n+\pi^+)\to d+\pi^0$ is negligible in the near threshold region (when the $\pi^0$ is produced in S- and P-states) - for the imaginary part of the corresponding threshold amplitudes. 
This result is obtained in a very general form, using only the symmetry 
properties of strong and 
electromagnetic interactions: the Pauli principle and conservation of 
P-parity and total 
angular momentum. It is a model independent result, and therefore it has to be 
verified by any model 
calculation. In particular, for numerical calculations, this would constitute a 
very important check that all contributions are properly treated and in particular double counting is avoided. 
The cancellation of RE can explain the absence of unitary cusp for the 
process $\gamma+d\to d+\pi^0$, at the threshold of processes 
$\gamma+d\to p+p+\pi^-$ and $\gamma+d\to n+n+\pi^+$.
Our result about RE changes the previous interpretation of the  experiment
\cite{Be98}, concerning the evaluation of the 
$E_{0+}(\gamma n\to n \pi^0)$-amplitude. The precise value of this amplitude is especially 
important in order to test 
the isotopic structure of $\gamma +N\to N+\pi$ and the predictions of 
ChPT \cite{VB96}. The 
interpretation of the data about $e+d\to e+d+\pi^0$ at large momentum transfer \cite{ETG98}, will have also to take into account this result. 

We have also shown that the same general arguments predict large RE, for 
another possible coherent process, the photoproduction of 'scalar' deuteron, 
$\gamma+d\to d_s+\pi^0$, with isotopic spin $I=1$. In this case the 
corresponding IA  amplitude is proportional to the difference of the 
$\gamma +p\to p+\pi^0$ and $\gamma +n\to n+\pi^0$ amplitudes.
\section{Aknowledgments}
We are pleased to thank J. M. Laget for many interesting discussions. One of us (M.P.R.) acknowledges the financial support of CEA (Saclay) and the warm hospitality of DAPNIA/SPhN, where this work was done.

\clearpage
\begin{table}

\begin{tabular}{|c|c|c|c|c|}
\hline\noalign{\smallskip}
&$E_d$ &$E_{0+}^{p \pi^0}$ &$E_{0+}^{n \pi^0}$&$E_{0+}^{n \pi^0}(IA)$\\
\noalign{\smallskip}\hline\noalign{\smallskip}
Exp~~ &$ \pm 1.45\pm 0.09$ ~\protect\cite{Be98} &$-1.31\pm 0.08$~ \protect\cite{Be97,Fu96} & {\large $^{2.5\pm 0.5}_{1.9\pm 0.3}$}~ \protect\cite{Ar88}  &{\large $^{~2.75 ~(+)}_{-0.15 ~(-)}$}\\
ChPT & $ -1.8\pm 0.2$ ~\protect\cite{Bea95,Bea97}& $-1.16$ ~\protect\cite{VB96} & $2.13$ \protect\cite{VB96} & \\
DR & $ - $&$-1.22$~\protect\cite{Ha97}& $1.19$~\protect\cite{Ha97}&\\
\noalign{\smallskip}\hline
\end{tabular}
\caption{Summary of the values of the elementary amplitudes from experiment and from model predictions, in units $10^{-3}/m_\pi$}
\label{tab1}
\end{table}

\begin{figure}

\begin{center}
\mbox{\epsfxsize=15.cm\leavevmode \epsffile{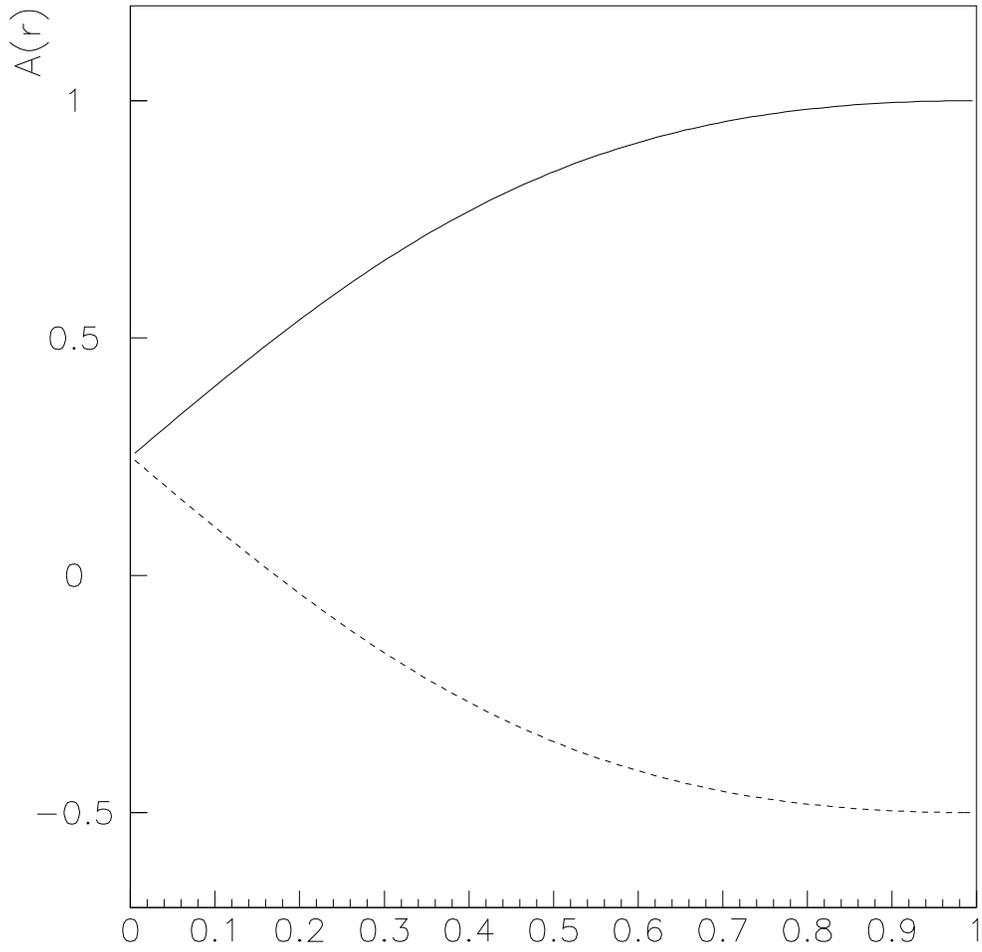}}
\end{center}
\vspace*{-1truecm}
\caption{Dependence of the tensor analyzing power on $r=|g_e|/|g_m|$.}
\label{fig:ratio}
\end{figure}

\begin{figure}
\begin{center}
\mbox{\epsfxsize=15.cm\leavevmode \epsffile{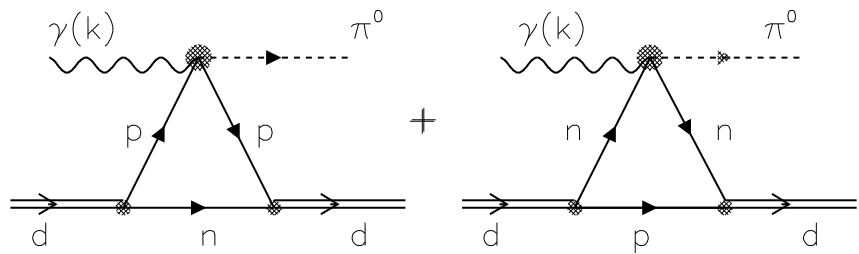}}
\end{center}
\vspace*{-6truecm}
\caption{Impulse approximation for $\gamma+d\to d+\pi^0$.}
\label{fig1}
\end{figure}

\begin{figure}

\begin{center}
\mbox{\epsfxsize=10.cm\leavevmode \epsffile{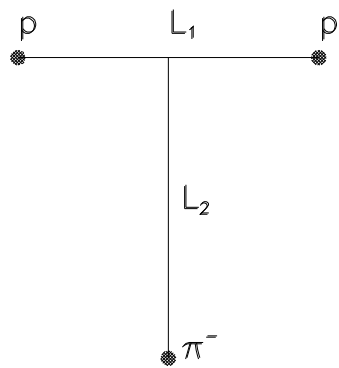}}
\end{center}
\caption{Definition of the orbital momentum $\ell_1$ and $\ell_2$ for the 
$pp\pi^-$-system.}
\label{fig2}
\end{figure}
\begin{figure}
\vspace*{-5truecm}
\begin{center}
\mbox{\epsfxsize=15.cm\leavevmode \epsffile{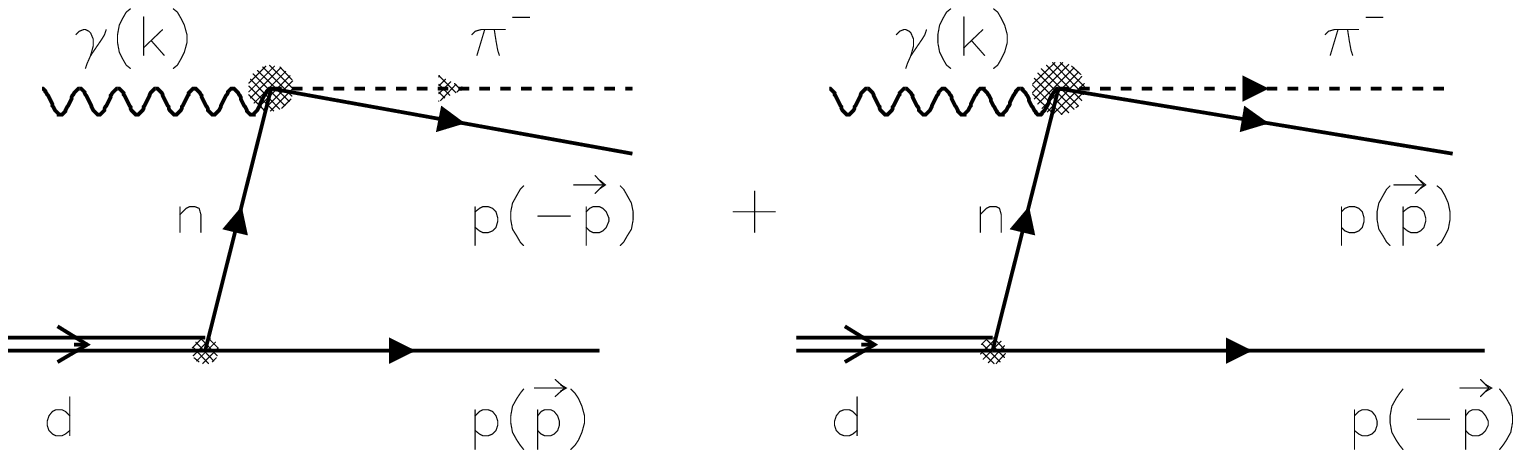}}
\end{center}
\vspace*{-5truecm}
\caption{Feynman diagram for  for $\gamma+d\to p p+\pi^-$.}
\label{fig3}
\end{figure}
%
\begin{figure}
\vspace*{-5truecm}
\begin{center}
\mbox{\epsfxsize=15.cm\leavevmode \epsffile{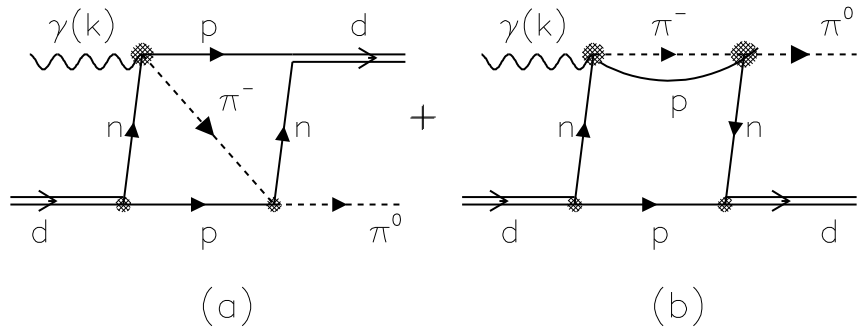}}
\end{center}
\vspace*{-5truecm}
\caption{Rescattering mechanism for $\gamma+d\to d+\pi^0$.}
\label{fig4}
\end{figure}
\newpage
\begin{figure}
\vspace*{-5truecm}
\begin{center}
\mbox{\epsfxsize=15.cm\leavevmode \epsffile{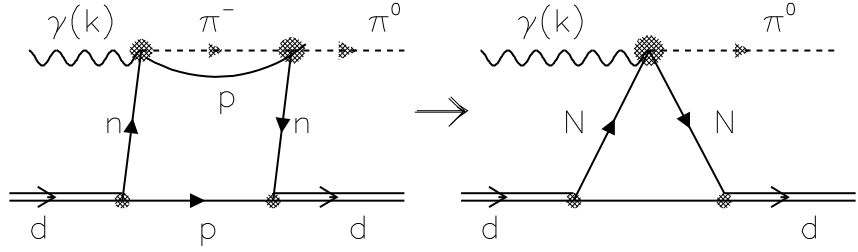}}
\end{center}
\vspace*{-6truecm}
\caption{Diagrams for possible double counting.}
\label{fig5}
\end{figure}
\begin{figure}
\vspace*{-3truecm}
\begin{center}
\mbox{\epsfxsize=17.cm\leavevmode \epsffile{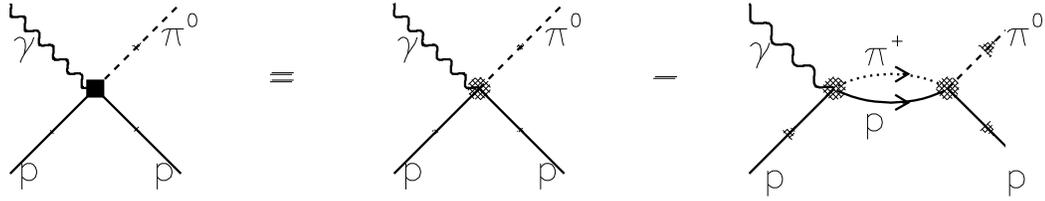}}
\end{center}
\vspace*{-12truecm}
\caption{Diagrammatic representation of the renormalized amplitude (solid 
square), whereas the solid circles represent the full amplitudes for 
$\gamma+p\to p+\pi^0$.}
\label{fig6}
\end{figure}
\begin{figure}
\vspace*{-5truecm}
\begin{center}
\mbox{\epsfxsize=15.cm\leavevmode \epsffile{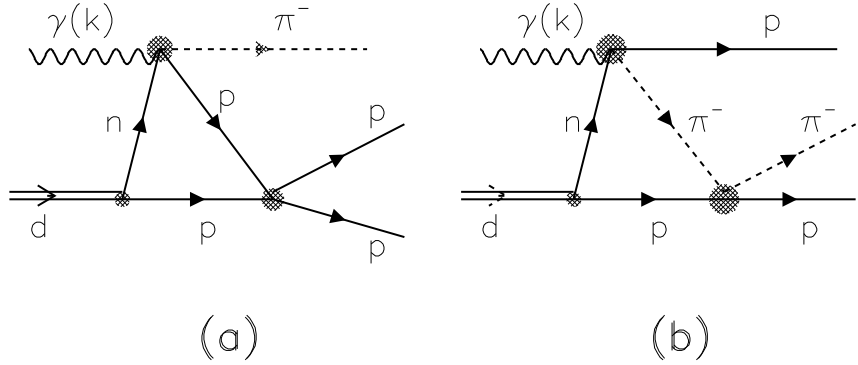}}
\end{center}
\vspace*{-4truecm}
\caption{Final state interaction in $\gamma+d\to \pi^-+p+p$.}
\label{fig7}
\end{figure}
\newpage
\begin{figure}
\vspace*{-6truecm}
\begin{center}
\mbox{\epsfxsize=15.cm\leavevmode \epsffile{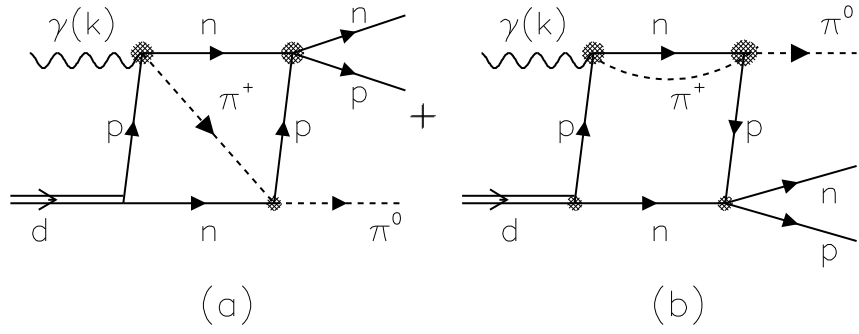}}
\end{center}
\vspace*{-5truecm}
\caption{Rescattering effects for the $ \gamma+d\to 
(p+n)+\pi^0$-process.}
\label{fig8}
\end{figure}
\begin{figure}
\vspace*{-5truecm}
\begin{center}
\mbox{\epsfxsize=15.cm\leavevmode \epsffile{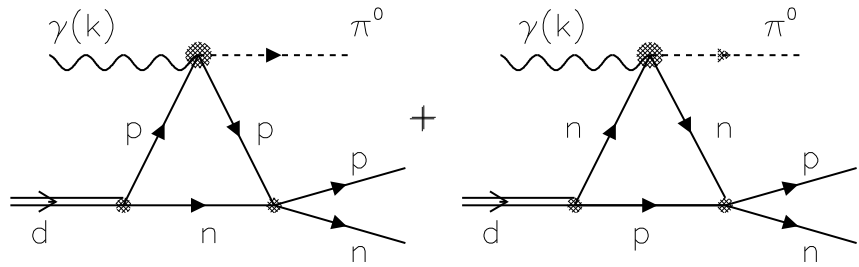}}
\end{center}
\vspace*{-5truecm}
\caption{Impulse approximation for the 
$\gamma+d\to (p+n)+\pi^0$-reaction}
\label{fig9}
\end{figure}

\end{document}